\RequirePackage{amsmath}
\pdfoutput=1

\documentclass[10pt]{iopart}

\usepackage{graphicx}

\usepackage{amsmath}
\usepackage{amsfonts}
\usepackage{mathtools}
\usepackage{amssymb}

\usepackage[colorlinks=true,linkcolor=blue,citecolor=blue,filecolor=cyan,urlcolor=blue,breaklinks=true]{hyperref}

\usepackage[utf8]{inputenc}

\usepackage{cite}

\newcommand{\avg}[1]{\langle#1\rangle}

\newcommand{\ket}[1]{|#1\rangle}
\newcommand{\bra}[1]{\langle#1|}

\renewcommand{\tr}[1]{\mathrm{tr}\{#1\}}

\newcommand{\kb}{k_\mathrm{B}}

\begin{document}
\title{Coherence-enhanced efficiency of feedback-driven quantum engines}
\author{Kay Brandner, Michael Bauer, Michael T. Schmid, Udo Seifert$^1$}
\ead{$^1$useifert@theo2.physik.uni-stuttgart.de}
\address{%
II. Institut für Theoretische Physik, Universität Stuttgart, 70550 Stuttgart, Germany
}%
\date{\today}

\begin{abstract}
A genuine feature of projective quantum measurements is that they
inevitably alter the mean energy of the observed system if the
measured quantity does not commute with the Hamiltonian.
Compared to the classical case, Jacobs proved that this additional energetic cost 
leads to a stronger bound on the work extractable after a single 
measurement from a system initially in thermal equilibrium [Phys. Rev. A 80, 012322 (2009)]. Here, we extend
this bound to a large class of feedback-driven quantum engines operating 
periodically and in finite time. The bound thus implies a natural definition 
for the efficiency of information to work conversion in such devices.
For a simple model consisting of a laser-driven two-level system, we
maximize the efficiency with respect to the observable whose 
measurement is used to control the feedback operations. We find
that the optimal observable typically does not commute with the 
Hamiltonian and hence would not be available in a classical two level 
system. This result reveals that periodic feedback engines operating in
the quantum realm can exploit quantum coherences to enhance efficiency. 
\end{abstract}

\section{Introduction}

Schrödinger's cat sums up one of the most striking and 
counter-intuitive features of quantum systems that is the 
ability to exist in coherent superpositions of states, which, in the 
classical world, would mutually exclude each other. 
While the conceptual ambiguities arising due to this phenomenon have
been highly debated in the early days of quantum mechanics, during the
last decades, it has been pointed out that quantum coherence might 
serve as valuable resource, especially for information processing. 
Among the first suggestions in this direction were the 
Brassard-Bennett protocol and the Deutsch–Jozsa algorithm promising 
respectively intrinsically eavesdrop-secure communication and an 
exponential speedup of computation by exploiting the quantum 
superposition principle \cite{niel10}.
Although theses schemes are of little practical use so far, they
reveal the enormous potential of quantum technologies, which 
nowadays becomes all the more significant due to recent experiments 
showing the accessibility of quantum effects even under ambient 
conditions \cite{dold13,wald14,engl15}. 

Information thermodynamics \cite{parr15,deff13,bara13d,horo14a} provides another, yet much less explored, 
area of research, which might benefit from the utilization of
quantum coherence. 
The development of this field was originally triggered by Maxwell's
famous thought experiment challenging the second law by invoking a
small intelligent being, which is able to separate the molecules of 
a gas in thermal equilibrium according to their velocity, thus establishing a spontaneous temperature gradient \cite{leff03}.
Building on Maxwell's idea, Szilárd invented a microscopic engine 
consisting of a single molecule confined in a container, which is in
contact with a thermal reservoir of constant temperature \cite{szil29}. 
An external agent might operate this setup by first dividing the 
container in two chambers, second, detecting the position of the 
molecule and, third, adiabatically expanding the chamber the molecule
was found in, thus extracting work from a single heat bath. 
Half a century after its discovery, this apparent contradiction with
the second law was resolved by Bennett \cite{benn82}, who showed that, due to
Landauer's principle \cite{land61}, the reduction of entropy associated with the 
measurement in the second step must be eventually compensated
when the external agent discards the gathered information from its 
memory, which can not be inexhaustible. 
Hence, effectively, the information acquired during the measurement 
is converted into work.
Meanwhile a fairly complete and experimentally confirmed theoretical 
framework exists \cite{parr15,toya10a,beru12} at least for classical systems, which, on a general
level,
provides precise extensions of the second law accounting for
information as a physical quantity thus relating it to traditional
thermodynamic variables such as entropy and work.

In the quantum realm, additional intricacies arise, which are not yet
fully explored
\cite{zure03,scul03,quan06,alla08,kim11a,jaco12,gelb13,yi13,stra13a,park13,funo13,horo14,gasp14,fren14,an15,gool15}. 
As a consequence of the superposition principle even the Hilbert 
space of a simple two-level system (TLS) contains infinitely many 
orthonormal pairs of realizable states, each of which is associated
with a specific observable, which, in principle, might be measured. 
Moreover, according to the projection postulate, a measurement will
typically alter the state of the system and thereby its mean energy. 
Therefore, in strong contrast to the classical case, a projective
quantum measurement is not only accompanied by a decrease in entropy
but also by an intrinsic change in internal energy, which must be 
taken into account for thermodynamic considerations. 
Jacobs argued that this energetic cost should be attributed to the 
external observer and derived an inequality, which incorporates it
in an upper bound on the work extractable from a quantum system in 
thermal equilibrium after a single measurement \cite{jaco09}. 
Here, we go one step further by relaxing the assumption on the 
initial state and allowing multiple measurements in finite 
intervals. 
Using a simple argument based on the first and the second law, we
show that Jacobs' bound holds whenever the probability to obtain a certain outcome does not change from one measurement to the next.
This result, in particular, implies a bound on the average work 
delivered by information driven quantum engines operating periodically
and in finite time.
Moreover, it provides a natural definition for the efficiency of 
such machines.

One of the first specific, fully quantum mechanical models for a 
measurement controlled device was proposed by Lloyd \cite{lloy97}.
In the spirit of Szilárd's pioneering work, he considers a single 
spin$-\frac{1}{2}$ system in contact with a thermal heat bath. 
An external controller can extract work in form of photons from 
this system by measuring the energy of the spin and applying a 
$\pi$-pulse at the Larmor frequency if the excited state is detected.
After the spin-flip, or, if initially the ground state was 
observed, the system is allowed to return to thermal equilibrium, 
before the procedure repeats. 
Lloyd demonstrates that his engine can completely convert the 
information acquired by the measurement into work. 
Furthermore, he argues that the efficiency of this process will 
inevitably decrease, due to decoherence effects, if any observable 
different from energy is used to determine the state of the system.
However, his reasoning strongly relies on the assumption that the spin
has relaxed to thermal equilibrium before any measurement, 
which, in fact, would require an infinite waiting time. 

In this work, by generalizing the setup described above, we show
that triggering a laser pulse by measuring an observable that does
not commute with the Hamiltonian of the system can enhance the efficiency if the model is operated in finite time. 
Specifically, we investigate a quantum-optical TLS, 
whose relaxation dynamics is modeled using a quantum master equation. 
After a projective von-Neumann measurement, the system is assumed
to be detached from the heat bath such that its time evolution 
during the laser pulse is governed by a  time-dependent Schr\"odinger
equation.
We note that such a separation of system and environment has recently
been argued to be realistic in the context of quantum heat engines
\cite{abah12,ross14}.
For our model, we analytically calculate the time-dependent density
matrix characterizing the system in the cyclic operation mode and
numerically determine the optimal observable to control the feedback
protocol as a function of the relaxation time and the spacing of the
energy levels.
Our findings show that exploiting quantum coherences can enhance the efficiency of information
engines beyond classically achievable values.

The paper is structured as follows. 
As our first main result, we derive a new bound on the average 
work output of quantum information engines in section 2.
In section 3, we introduce a specific model for such a machine
and solve its dynamics.
Section 4 is devoted to the optimization of its efficiency. 
We conclude in section 5.

\section{Bound on work for cyclic quantum information engines}
\renewcommand{\r}{\rho}
\newcommand{\ri}{\r_{{\rm ini}}}
\newcommand{\V}{\mathcal{V}}
\renewcommand{\S}{S_{{\rm sys}}}
\renewcommand{\tr}[1]{{{\rm tr}}\left\{#1\right\}}

We begin this section by introducing a general scheme for a cyclic 
quantum information engine.
To this end, we consider a finite quantum system with Hamiltonian
$H$, which is in contact with a heat bath of temperature
$T$ and whose density matrix is initially given by 
$\r_{{\rm ini}}$. 
This setup is now operated by an external agent in two steps. 
First, an instantaneous projective measurement of the observable $A$ is carried
out, which yields the outcome $a_m$ and leaves the system in the state
\begin{equation}
\r_m\equiv\ket{\psi_m}\bra{\psi_m}
\end{equation}
with probability 
\begin{equation}
p_m^{(0)}\equiv \bra{\psi_m}\r_{{\rm ini}}\ket{\psi_m}.
\end{equation}
Here, the $a_m$ are the eigenvalues of $A$, which we assume to be 
non-degenerate, and $\ket{\psi_m}$ denotes the normalized eigenvector
of $A$ corresponding to $a_m$. 
Second, to convert the acquired information into useful work, 
a control operation is applied to the system, which is conditioned on
the result of the preceding measurement and leads to the evolved density
matrix 
\begin{equation}\label{COp}
\tilde{\r}_m\equiv \V_m[\r_m],
\end{equation}
where $\V_m$, in principle, can be any positive, trace preserving map
\cite{niel10}. 
Practically, such an operation can be realized by intermediately manipulating the
Hamiltonian of the system or its coupling to the heat bath over a 
certain time interval.

\begin{figure}
 \includegraphics{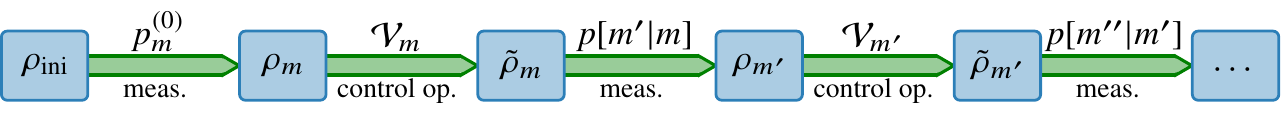}
\caption{Flow chart illustration of the operation principle of a general quantum information engine. Alternately, the observable $A$ is measured and the control operation $\V_m$ conditioned on the measurement outcome $a_m$ is applied to the system. Symbols are explained in the main text.
}
\label{fig:flow_diag}
\end{figure}

The agent now iterates the sequence of steps one and two, where,
in the $i$\textsuperscript{th} operation cycle, the measurement 
outcome $a_{m'}$ is obtained with probability $p_{m'}^{(i)}$. 
It is readily seen that these quantities fulfill the recursion 
relation
\begin{equation}
p_{m'}^{(i)}= \sum_{m} p[m'|m]p_{m}^{(i-1)}
\end{equation}
with the conditional probability
\begin{equation}\label{TransProb}
p[m'|m] =\bra{\psi_{m'}}\tilde{\r}_{m}\ket{\psi_{m'}}   
        =\bra{\psi_{m'}}\V_{m}\big[\ket{\psi_{m}}
         \bra{\psi_{m}}\big]\ket{\psi_{m'}},
\end{equation}
since the initial density matrix of each cycle is the result of the
control operation applied in the foregoing one, as shown in figure \ref{fig:flow_diag}.
Moreover, since the transition probability \eqref{TransProb} does not
depend on the cycle index $i$ but rather is fully determined by the
control operation $\V_m$ and the observable $A$, after sufficiently
many iterations a stationary distribution $q_m\equiv
\lim_{i\rightarrow\infty}p_m^{(i)}$ satisfying 
\begin{equation}\label{SS}
q_{m'} = \sum_m p[m'|m]q_{m}
\end{equation}
will be approached.
Once this steady state is reached, the system works as a periodic
information engine. 

For a thermodynamic analysis of the scheme outlined above, we 
have to calculate the changes in internal energy $E[\r]\equiv
\tr{H\r}$ and entropy $\S[\r]\equiv-\kb \tr{\r\ln\r}$ of the system
associated with the steps one and two, where $\kb$ is Boltzmann's constant. 
Considering an operation cycle with initial density matrix 
$\tilde{\r}_{m}$ and measurement outcome $a_{m'}$, we find
\begin{align}
\Delta E^{{\rm meas}}(m',m) &=  E[\r_{m'}]- E[\tilde{\r}_{m}] 
                 =  \bra{\psi_{m'}}H\ket{\psi_{m'}} -\tr{H\tilde{\r}_{m}},
                 \label{DEMeas}\\
\Delta\S^{{\rm meas}}(m',m) &=  \S[\r_{m'}]- \S[\tilde{\r}_{m}]
                 =  \kb \tr{\tilde{\r}_{m}\ln\tilde{\r}_{m}}
                 \label{DSMeas}
\end{align}
for the measurement and 
\begin{align}
\Delta E^{{\rm con}}({m'})  &= E[\tilde{\r}_{m'}]- E[\r_{m'}] 
             = \tr{H\tilde{\r}_{m'}}-\bra{\psi_{m'}}H\ket{\psi_{m'}},\\
\Delta\S^{{\rm con}}({m'}) &= \S[\tilde{\r}_{m'}]-\S[\r_{m'}]
             = -\kb \tr{\tilde{\r}_{m'}\ln\tilde{\r}_{m'}}
\end{align}
for the control operation, where we used $\S[\r_{m'}]=0$ due to $\r_{m'}$ 
representing a pure state. 
Since the total entropy production during the control step
\begin{align}
 \Delta
S^{{\rm con}}_{{\rm tot}}({m'})= \Delta \S^{{\rm con}}({m'})
+\Delta S^{{\rm con}}_{{\rm bath}}({m'})\geq 0
\end{align}
must be nonnegative by virtue
of the second law, it follows that the change in entropy of the heat
bath $\Delta S^{{\rm con}}_{{\rm bath}}({m'})$ is bounded from
below by $-\Delta \S^{{\rm con}}({m'})$ and thus the heat taken up by the system $Q({m'})= -T\Delta S^{{\rm con}}_{{\rm bath}}({m'}) $
during the control operation $\V_{m'}$ is bounded from above by
$T\Delta \S^{{\rm con}}({m'})$.
Consequently, the first law
\begin{equation}
\Delta E^{{\rm con}}({m'}) = Q({m'}) - W({m'})
\end{equation}
implies the bound 
\begin{equation}\label{BoundW}
W({m'})\leq T\Delta \S^{{\rm con}}({m'})-\Delta E^{{\rm con}}({m'})
\end{equation}
on the work $W({m'})$ the agent can extract from the system using the
operation $\V_{m'}$.

Since the measurement outcome $a_m$ occurs with probability $q_m$ in
the steady state, \eqref{BoundW} yields the bound 
\begin{align}
\langle W\rangle &\equiv \sum_{m} q_m W(m)\\
                 &\leq \sum_{m} q_m\Big(
                   -\kb T\tr{\tilde{\r}_m\ln\tilde{\r}_m}
                   -\tr{H\tilde{\r}_m}
                   +\bra{\psi_m}H\ket{\psi_m}\Big)             
                   \label{BWork2}     
\end{align}
on the average work extracted per operation cycle.
Furthermore, the average energetic cost and entropy reduction per 
cycle associated with the measurement read 
\begin{align}
\langle\Delta E^{{\rm meas}}\rangle &
\equiv\sum_{m',m} p(m',m)\Delta E^{{\rm meas}}(m',m)
\label{DEMeasAv}
\end{align}
and
\begin{align}
\langle\Delta \S^{{\rm meas}}\rangle &
\equiv\sum_{m',m} p(m',m)\Delta \S^{{\rm meas}}(m',m),
\label{DSMeasAv}
\end{align}
respectively. 
Here, $p(m',m)\equiv p[m'|m]q_{m}$ is the probability to measure
$a_{m'}$ and $a_m$ in two consecutive operation cycles.
Inserting \eqref{DEMeas} and \eqref{DSMeas} into \eqref{DEMeasAv} 
and \eqref{DSMeasAv} and using the steady state condition \eqref{SS}
as well as the sum rule 
\begin{equation}\label{SumRule}
\sum_{m'} p[m'|m] = 1
\end{equation}
expressing probability conservation yields
\begin{align}
\langle\Delta E^{{\rm meas}}\rangle & = 
\sum_m q_m\left(\bra{\psi_m}H\ket{\psi_m}-\tr{H\tilde{\r}_m}\right),
\label{DEMeasAv2}\\
\langle\Delta \S^{{\rm meas}}\rangle & = 
\sum_m q_m\kb \tr{\tilde{\r}_m\ln\tilde{\r}_m}.
\label{DSMeasAv2}
\end{align}
By comparing \eqref{DEMeasAv2} and \eqref{DSMeasAv2} with
\eqref{BWork2}, we obtain the bound
\begin{equation}\label{WBound}
\langle W\rangle \leq - T\langle\Delta\S^{{\rm meas}}\rangle +
\langle\Delta E^{{\rm meas}}\rangle.
\end{equation}
This inequality, which constitutes our first main result, provides a universal upper bound on the average work 
extractable per operation cycle in terms of quantities that 
are related
to the measurement process only. 
It generalizes similar results obtained in \cite{jaco09,saga08,hase10,erez12} for single stroke operations.
Following the arguments of Jacobs \cite{jaco09}, we consider the energetic cost of the 
measurement $\langle \Delta E^{{\rm meas}}\rangle$ as work input 
provided by the measurement apparatus and thus infer from
\eqref{WBound} the natural definition
\begin{equation}\label{Eff}
\eta\equiv 
\frac{\langle W\rangle}{\langle \Delta E^{{\rm meas}}\rangle-
T\langle\Delta \S^{{\rm meas}}\rangle}\leq 1
\end{equation}
for the efficiency, at which information is converted to work in cyclic quantum engines.
We note that, while $-\langle\Delta\S^{{\rm meas}}\rangle$ is readily
seen to be always nonnegative, in contrast to the setup considered
in \cite{jaco09}, $\langle\Delta E^{{\rm meas}}\rangle$ can, in principle, become
negative, since, for finite cycle times, the system will typically not
be in thermal equilibrium before the measurement is performed. 
Moreover, the quantity $\langle\Delta E^{{\rm meas}}\rangle$ is of 
pure quantum origin and vanishes in the quasi-classical situation, 
where the observable $A$ commutes with the Hamiltonian of the system
$H$.

\section{Quantum optical model}
\renewcommand{\t}{\theta}
\renewcommand{\L}{\mathcal{L}}
\newcommand{\G}{\Gamma}

As an application of the general theory discussed so far, we propose a generalization of a paradigmatic model for 
a quantum information engine originally invented by Lloyd \cite{lloy97} and
analyze its thermodynamic properties. 
Specifically, we consider an optical TLS with Hamiltonian
\begin{equation}
H=\frac{\hbar\omega_0}{2}\left(\ket{e}\bra{e} - \ket{g}\bra{g}\right),
\end{equation}
where $\hbar\omega_0>0$ is the energetic spacing between the ground
state $\ket{g}$ and the excited state $\ket{e}$.
The external agent measures the observable $A(\t)$ $(0\leq \t \leq \pi/2)$ with eigenvalues
$a_\pm=\pm 1$ and corresponding eigenvectors 
\begin{align}
\ket{\psi_+(\t)} &\equiv \ket{\psi_+}\equiv \cos\frac{\t}{2}\ket{e}
                        +\sin\frac{\t}{2}\ket{g},\\
\ket{\psi_-(\t)} &\equiv \ket{\psi_-}\equiv\sin\frac{\t}{2}\ket{e}
                        -\cos\frac{\t}{2}\ket{g},
\end{align}
which reduce to the eigenstates of $H$ for $\t=0$. 
In order to extract work in form of photons,
after a measurement of the state
$\ket{\psi_+}$, the system is
detached from the heat bath and a coherent laser pulse on resonance is applied for an interval $t_F$.
After a measurement of $\ket{\psi_-}$, the system is kept in contact with the thermal environment
for a time $t_R$ without any action of the agent to allow the 
absorption of additional heat before the next measurement is carried
out.

\begin{figure}
 \centering
\includegraphics{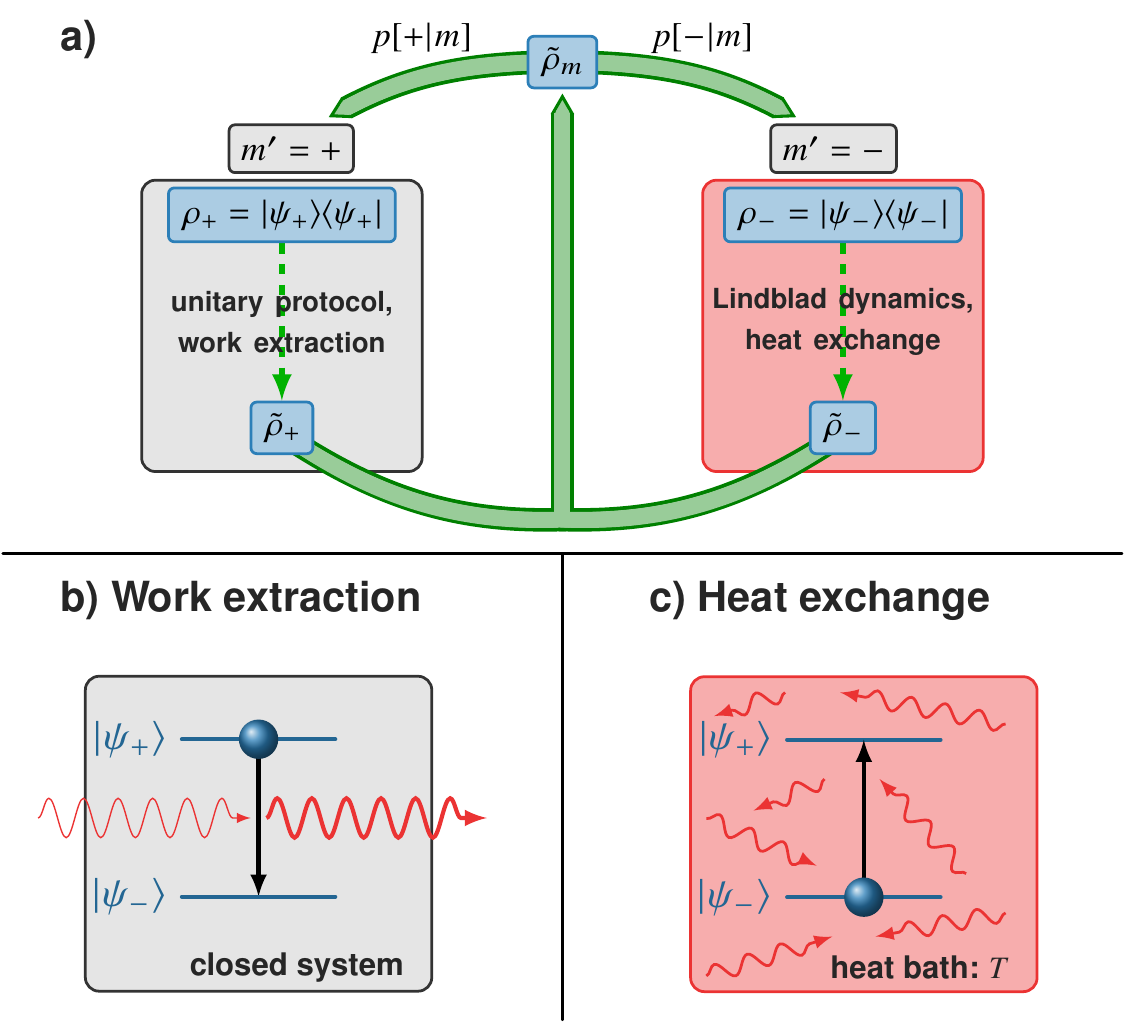}
\caption{Scheme of the quantum optical TLS as 
information engine (a). 
At the beginning of each operation cycle,
the state of the system is inferred by the agent via a measurement of
the observable $A(\t)$. 
If the outcome of this measurement is $a_+$, the internal energy of 
the TLS is used to coherently amplify an externally generated
laser pulse (b). 
If the outcome is $a_-$, the susceptibility of the system
for further energy uptake is exploited to extract heat from the 
environment (c). 
In any case, the density matrix at the end of the operation cycle
serves as initial state for the subsequent one. 
For further explanations of the symbols, see main text.}
\label{fig:TLS_scheme}
\end{figure}

For a quantitative description of this procedure, which is summarized 
in figure \ref{fig:TLS_scheme},
we need to specify the control operations $\V_\pm$.
During the interaction ($0\leq \tau \leq t_F$) with the laser pulse, the density matrix
$\r(\tau)$ of the system evolves unitarily according to the 
Liouville-von Neumann equation 
\begin{equation}\label{LEq}
\partial_\tau\r(\tau) = -\frac{i}{\hbar}[H_F(\tau),\r(\tau)] 
                     \equiv \mathcal{L}_+(\tau)\r(\tau),
\end{equation} 
where, on the semiclassical level and within the rotating wave
approximation, the time-dependent Hamiltonian is given by
\begin{equation}
H_F(\tau)\equiv H + \frac{\hbar\Omega}{2}\left(
e^{-i(\omega_0\tau-\phi)}\ket{e}\bra{g} 
+ e^{i(\omega_0\tau-\phi)}\ket{g}\bra{e}\right)
\end{equation}
with real Rabi frequency $\Omega>0$ and $0\leq\phi < 2\pi$ being
the phase of the dipole matrix element \cite{scul08}.
To describe the interaction of the TLS with the heat bath during $0\leq \tau \leq t_R$, we use the
quantum optical master equation \cite{breu07}
\begin{align}
\partial_\tau\r(\tau) & = -\frac{i}{\hbar}[H,\r(\tau)]
+ \gamma N\left(L\r(\tau)L^\dagger 
                -\frac{1}{2}L^\dagger L\r(\tau)
                -\frac{1}{2}\r(\tau)L^\dagger L\right)\nonumber\\
&\hspace{2.23cm} + \gamma (N+1)\left(L^\dagger\r(\tau)L 
                -\frac{1}{2}LL^\dagger\r(\tau)
                -\frac{1}{2}\r(\tau)LL^\dagger\right)\nonumber\\
&\equiv \L_-\r(\tau),
\label{MEq}
\end{align}
where $L\equiv \ket{e}\bra{g}$ is a Lindblad operator, $N\equiv 1/(\exp[\hbar\omega_0/
(k_B T)]-1)$  denotes the Planck distribution evaluated for the level
spacing $\hbar\omega_0$ and $\gamma>0$ is a damping rate quantifying
the coupling strength between the TLS and the thermal reservoir.
This time evolution equation, which is of Lindblad form and therefore
preserves trace and positivity of the density matrix, can be derived
from a microscopic model in the weak coupling limit, where the role of
the heat bath is played by the thermal radiation field, for details, 
see \cite{breu07}.
Such master equations are a well established method for the description of
open quantum systems, which has previously lead to substantial
insights in the context of quantum heat engines, see for example 
\cite{geva94,geva95}.
In terms of the super operators $\L_\pm$, the control operations admit
the formal representations
\begin{equation}
\V_+[\r] =\overrightarrow{\mathcal{T}}e^{\int_0^{t_F}\!\!\! d\tau\;\L_+(\tau)}\r
\quad\text{and}\quad
\V_-[\r] = e^{\L_- t_R}\r,
\end{equation}
where $\overrightarrow{\mathcal{T}}$ indicates
time ordering. 
Solving the equations \eqref{LEq} and \eqref{MEq} for a general
initial condition yields the explicit expressions \cite{scul08}
\begin{align}
\V_+[\r] & = U\r U^\dagger  \quad\text{with}\quad\nonumber\\
U & = \cos\frac{\Omega t_F}{2}\left(
e^{i\omega_0t_F/2}\ket{g}\bra{g}
+ e^{-i\omega_0t_F/2}\ket{e}\bra{e}\right)\nonumber\\
&\qquad -i\sin\frac{\Omega t_F}{2}\left(
e^{i\left(\omega_0t_F/2-\phi\right)}L^\dagger
+ e^{-i\left(\omega_0t_F/2-\phi\right)}L\right)
\label{COPp}
\end{align}
and \cite{naka06}
\begin{align}
\V_-[\r] = e^{\L_- t_R}\r & =
\frac{1}{4}\left(1+e^{-\G t_R}+2e^{-\G t_R/2}
\cos\omega_0 t_R\right)\r\nonumber\\
&+\frac{1}{4}\left(1+e^{-\G t_R}-2e^{-\G t_R/2}
\cos\omega_0 t_R\right)L_0\r L_0\nonumber\\
&-\frac{1}{4}\left(\frac{\gamma}{\G}(1-e^{-\G t_R})
 -2ie^{-\G t_R/2}\sin\omega_0 t_R\right)\r L_0\nonumber\\
&-\frac{1}{4}\left(\frac{\gamma}{\G}(1-e^{-\G t_R})
 +2ie^{-\G t_R/2}\sin\omega_0 t_R\right)L_0\r\nonumber\\
&+(1-e^{-\G t_R})\left(\frac{\gamma (N+1)}{\G}L^\dagger\r L 
                      +\frac{\gamma N}{\G} L\r L^\dagger\right),
\label{COPm}
\end{align}
where we introduced the abbreviations 
$L_0\equiv\ket{e}\bra{e}-\ket{g}\bra{g}$ and $\G\equiv\gamma (2N +1)$.

The work extracted within an operation cycle with measurement outcome
$a_+$ can be determined form the first law 
\begin{equation}\label{Work}
W(+) = -\Delta E^{{\rm con}}(+) = E[\r_+] - E[\tilde{\r}_+]
     = E[\r_+] - E[\V_+[\r_+]],
\end{equation}
since the TLS is decoupled from the environment and thus no heat is
exchanged during the control operation $\V_+$. 
Inserting $\r_+=\ket{\psi_+}\bra{\psi_+}$ and \eqref{COPp} into
\eqref{Work} gives 
\begin{equation}\label{Work2}
W(+) = \frac{\hbar\omega_0}{2}
\Big(\cos\t\left(1-\cos\Omega t_F\right)
-\sin\t\sin\phi\sin\Omega t_F\Big).
\end{equation}
To keep the subsequent analysis as simple as possible, from here 
onwards, we fix the pulse duration $t_F$ and the dipole phase $\phi$ 
such that \eqref{Work2} assumes the maximal value
\begin{equation}
W(+) = \hbar\omega_0 \cos ^2 \frac{\t}{2}
\end{equation}
with respect to these parameters, i.e., we put
\begin{equation}
t_F  = \frac{\t+\pi}{\Omega}
\qquad\text{and}\qquad
\phi = \frac{\pi}{2}.
\end{equation}
This choice ensures that the TLS ends up in the ground state after
the laser pulse, i.e., $\tilde{\r}_+=\ket{g}\bra{g}$. 
Furthermore, it leads to the fairly simple expressions
\begin{align}
p[+|+] &= \sin^2\frac{\t}{2}
\end{align}
and
\begin{align}
p[+|-] &=
\frac{1}{2}\left(1-\frac{\cos \t}{2N+1}-e^{-\G t_R/2}\sin^2\t\cos\omega_0 t_R
  -e^{-\G t_R}\cos\t \left[\cos \t-\frac{1}{2N+1} \right]                
                          \right)\label{CondProb}
\end{align}
for the conditional probabilities defined in \eqref{TransProb}. 
Since $p[-|+]$ and $p[-|-]$ are determined by the sum rules
\eqref{SumRule}, the steady state probabilities $q_\pm$ can now be
obtained from the fixed point condition \eqref{SS}. 
Specifically, we find 
\begin{equation}
q_+ = \frac{p[+|-]}{1-p[+|+]+p[+|-]} = 1-q_-.
\end{equation}

We are now ready to calculate the quantities entering the efficiency
\eqref{Eff}. 
First, the average work per cycle reads
\begin{equation}\label{EtaW}
\langle W\rangle = q_+ W(+),
\end{equation}
since no contribution arises from operation cycles with measurement
outcome $a_-$. 
Second, the average energy spent on the measurement \eqref{DEMeasAv2}
becomes
\begin{align}
\langle \Delta E^{{\rm meas}}\rangle 
& =\sum_{m=\pm}q_m\Big(E[\r_m]- E\big[\V_m[\r_m]\big]\Big)\nonumber\\
& = \langle W\rangle + q_-\frac{\hbar\omega_0}{2}
    \left(1- e^{-\G t_R}\right)\left(\frac{1}{2N+1}-\cos\t\right)
    \label{EtaDEMeas}
\end{align}
upon using $\r_\pm=\ket{\psi_\pm}\bra{\psi_\pm}$ and the expressions
\eqref{COPp} and \eqref{COPm} for the control operations. 
Third, since $\tilde{\r}_+$ represents a pure state due to the control
operation $\V_+$ being unitary, the average entropy reduction in the
system associated with the measurement \eqref{DSMeasAv2} arises only
from cycles with initial state $\tilde{\r}_-$. 
After some algebra again using \eqref{COPm}, we thus obtain
\begin{align}
\langle \Delta \S^{{\rm meas}}\rangle &= q_-\kb \tr{\V_-[\r_-]\ln\V[\r_-]}
\nonumber\\
&= \frac{q_-}{2}\kb \left(\ln D +\sqrt{1-4D}\ln\left(\frac{1+\sqrt{1-4D}
}{1-\sqrt{1-4D}}\right)\right)
\qquad\text{with}\label{EtaDSMeas}\\
D&\equiv \frac{1}{4}\left(1- \left(e^{-\G t_R}\cos\t
-\frac{e^{-\G t_R}-1}{2N+1}\right)^2 - e^{-\G t_R}\sin^2\t\right).
\label{EtaDDef}
\end{align}
Using the expressions \eqref{EtaW}-\eqref{EtaDDef}, the efficiency 
$\eta$ of this quantum optical information engine can
be evaluated for any complete set of parameters 
comprising the level spacing $\hbar\omega_0$, the temperature of the
heat bath $T$, the angle $\t$, the 
damping rate $\gamma$ and the relaxation time $t_R$.

\section{Quasi-classical vs. coherence-enhanced regime}
\newcommand{\g}{\gamma}
\renewcommand{\o}{\omega}

\begin{figure}
 \centering
\includegraphics[width=0.99\textwidth]{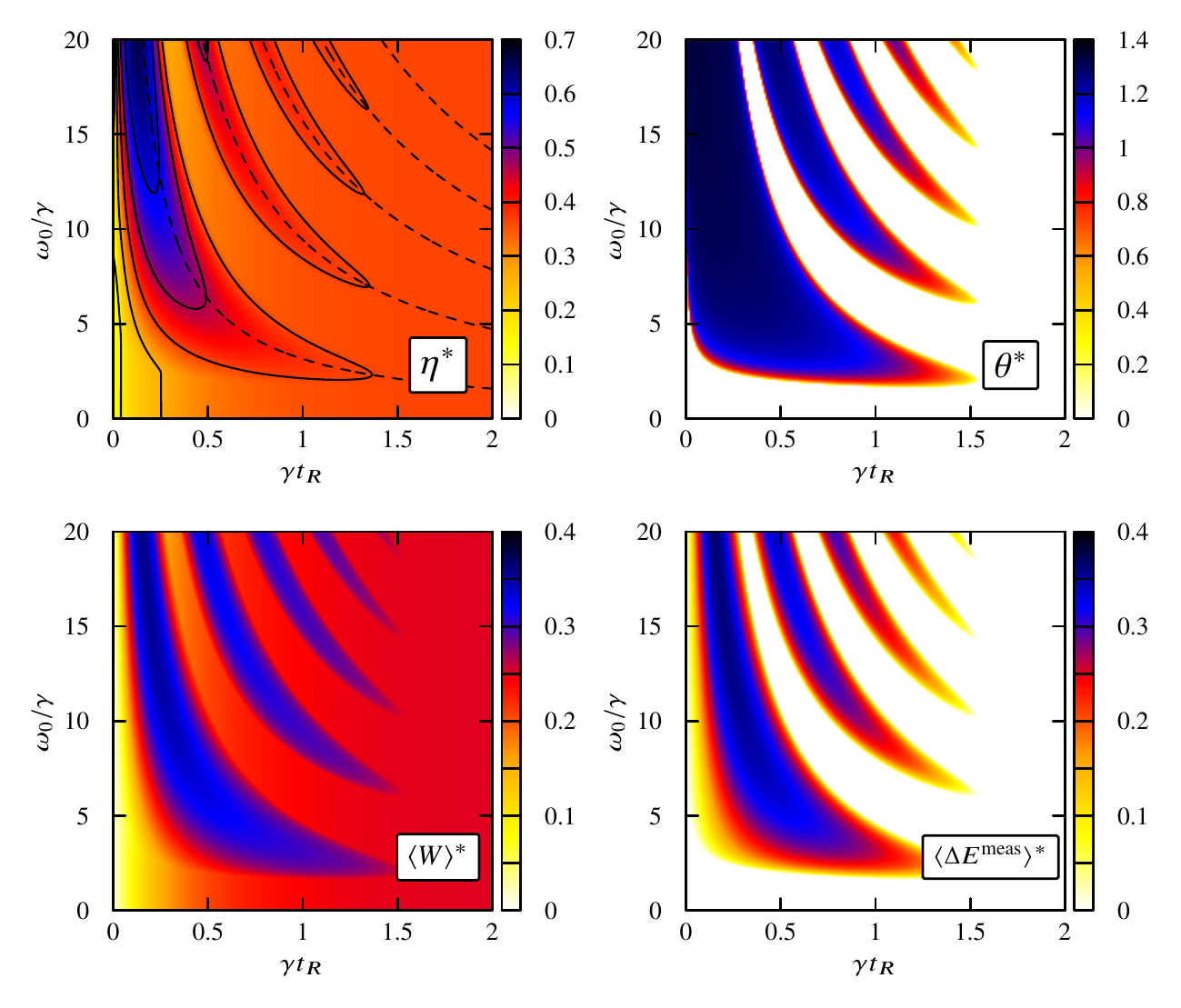}
\caption{Benchmark parameters for the performance of the optimized 
quantum optical TLS information engine as functions of the 
dimensionless parameters $\g t_R$ and $\o_0/\g$. 
The upper panel shows the maximum efficiency $\eta^\ast$ on the left 
and the corresponding optimal angle $\t^\ast$ on the right.
Along the dashed lines, the condition \eqref{MaxCCond} is fulfilled. 
The solid lines, which have a spacing of $0.1$ and constant offset 
$0.1625$, were introduced for graphical purposes. 
In the lower panel, the work output $\langle W\rangle^*$ (left)
and the average energy input required per measurement $\avg{\Delta 
E^{\rm meas}}^*$ (right) is plotted in units of $\hbar\omega_0$.}
\label{fig:eta_theta}
\end{figure}

In this section, we focus on the question whether coherences, i.e., a
choice $\t\neq 0$ for the basis, in which the measurement is 
performed, can enhance the efficiency $\eta$.
In order to reduce the number of free parameters, we choose from now
on the temperature such that 
\begin{align}
 x\equiv \hbar \omega_0/\kb T=\ln 2,
\end{align}
leading to $N=1$.
We then find by inspection that $\eta$ depends only on $\t$ and the
two dimensionless parameters $\gamma t_R$ and $\omega_0/\gamma$.
A numerical optimization procedure yields the maximal efficiency
$\eta^*$ and optimal angle $\t^*$, which are both shown in upper
panels of figure \ref{fig:eta_theta}.
These plots exhibit two qualitatively different regimes separated by
$\gamma t_R\simeq 1.5$.

First, for $\g t_R\gtrsim 1.5$, we recover the quasi-classical regime 
originally considered by Lloyd \cite{lloy97}, within which the TLS can
relax to thermal equilibrium in each operation cycle with measurement
outcome $a_-$.
As argued in \cite{lloy97}, the largest efficiency can then be
obtained for $\t=0$.
Consistently, we observe that $\t^\ast$ is effectively zero in this
regime and, independent of $\omega_0/\gamma$, the optimal efficiency
settles at the constant value
\begin{align}
\eta^\ast & \approx \frac{x}{(1+e^x)\ln(1+e^x)-x e^x}\simeq 0.36.
\label{EffLRT}
\end{align}

Second, in the coherent regime $\gamma t_R\lesssim 1.5$, we find a
characteristic oscillatory pattern, which can be traced back to the 
structure of the conditional probability \eqref{CondProb}.
The crucial role of this quantity, which is, in fact, the only 
ingredient of the efficiency depending on $\omega_0/\gamma$, can be 
explained by the following argument.
If the TLS is found in the state $\ket{\psi_+}$ after being in contact
with the thermal environment for the time $t_R$, the heat absorbed 
during this period together with the energy invested for the 
measurement can be converted into work by applying a laser pulse.
A measurement indicating the state $\ket{\psi_-}$, however, leads 
to another relaxation cycle, within which no work can be extracted and
the previously gained information is inevitably wasted. 
Consequently, the efficiency, at which acquired information is
converted into work, can be expected to increase as the frequency of
such idle cycles decreases.
For $\gamma t_R\lesssim 1.5$, the corresponding probability
\begin{align}
p[-|-] & = 1- p[+|-]\nonumber\\
       & = \frac{1}{2}\left( 1+\frac{\cos\t}{3}+e^{-3\g t_R/2}\sin^2\t
       \cos \o_0 t_R+ e^{- 3\g t_R}\cos\t\left[\cos\t -\frac{1}{3}
       \right]\right)
     \label{CondProbIC}
\end{align}
can be substantially reduced by the contribution proportional to 
$e^{-3\g t_R/2}$, which arises solely due to quantum coherences and
vanishes for $\t=0$. 
This effect becomes most pronounced for $\t=\pi/2$ and 
\begin{equation}\label{MaxCCond}
\o_0 t_R = (\o_0/\g)\cdot (\g t_R) = (2n+1)\pi, 
\qquad (n=0,1,2,\dots).
\end{equation}
Accordingly, the hyperbolas \eqref{MaxCCond} are in good agreement
with the local maxima of the efficiency $\eta^\ast$ in the 
$(\g t_R$,$\o_0/\g)$-plane and $\theta^\ast$ comes close to $\pi/2$ in
their vicinity.
The deviations from this pattern for $\g t_R\lesssim 0.5$ can be 
explained by the remaining terms in \eqref{CondProbIC}, which come
with a prefactor $e^{-3\g t_R}$ and thus give a non-negligible 
contribution only in these regions.
Most importantly, in this regime, we find as our second main result 
that the efficiency is enhanced by exploiting coherences.
Specifically, it can overcome the quasi-classical value \eqref{EffLRT}
and even approach its upper bound $1$ in the limit $\omega_0/\gamma
\rightarrow\infty$.  

The average work output for the optimal angle $\t^\ast$, $\langle
W\rangle^\ast$, is plotted in the lower panel of figure 
\ref{fig:eta_theta}.
Clearly, this quantity features the same characteristic dependence
on $\g t_R$ and $\o_0/\g$ as the maximized efficiency $\eta^\ast$. 
Like $\eta^\ast$, the average work $\langle W\rangle^\ast$ exceeds its 
quasi-classical limit 
\begin{equation}
\lim_{\g t_R\rightarrow\infty} \langle W \rangle^\ast = \hbar \omega_0/4
\end{equation}
in the coherence-enhanced regime and becomes maximal in the same 
range of parameters like $\eta^\ast$, i.e., in the vicinity of the
hyperbolas \eqref{MaxCCond}.

Finally, we consider the average energetic
cost per measurement $\langle\Delta E^{{\rm meas}}\rangle^\ast$. 
This additional input is inevitably necessary for the exploitation of
quantum coherence and therefore becomes non-negligible whenever 
$\t^\ast$ significantly deviates from  $0$, hence, in particular, in the 
regions of the parameter space, where our numerical procedure reveals 
$\eta^\ast$ to be large. 
Consequently, in the range of high efficiencies, the input of the 
device is mainly delivered by the measurement apparatus rather than
the heat bath. 
This result underlines the crucial role of the measurement process
in the quantum realm, which, besides delivering information, 
can alter the state of the system and thus bears the character of 
an additional control operation.

\section{Conclusion}

In this paper, we have derived a universal upper bound on the
average work output delivered in finite time by cyclically operating
quantum information engines, which takes into account the energetic 
cost intrinsically associated with quantum measurements.
This bound provides a benchmark for the performance of quantum 
mechanical information-to-energy converters, which, in contrast to 
their classical counterparts, see for example \cite{baue12,baue14, 
alla09,abre11,horo11a,mand12,horo12a,bara13a,sand14}, can exploit 
the superposition principle and thus might be able to overcome 
classical limitations.

We have explicitly investigated the benefit of quantum coherences in 
the second part of the paper by considering a specific model 
consisting of a quantum optical TLS, which, conditioned on the outcome
of a projective measurement, is repeatedly either coupled to a heat 
bath or used to amplify a coherent laser pulse.
In the regime of long relaxation times, this setup corresponds to a
model originally proposed by Lloyd, whose properties are reproduced
qualitatively in our analysis within this limit.
We emphasize, however, that the definition of efficiency used in
\cite{lloy97} is different from ours, since it explicitly refers to
Landauer's principle by invoking the minimal heat that must be
dissipated in a second heat bath of different temperature to achieve
the entropy production necessary to reset the memory of the external
agent.
Viewed in this way, the model acts effectively as a heat engine, whose
efficiency is bounded by the Carnot value.
In our approach, we consider the system as an information engine and
define its efficiency in terms of quantities directly associated
with the system and the measurement process, leaving aside
how the agent eventually erases the gathered information.

In the coherent regime, which is characterized by short cycle times,
we find that utilizing a non-classical observable $A(\t)$, whose
eigenstates are coherent superpositions of the energy eigenstates,
can enhance the performance significantly.
Remarkably, it turns out that both, efficiency and average work output
per cycle, can be substantially increased if the relaxation time is
properly adjusted to the level spacing.
Since, in the corresponding regions of the parameter space, the
optimal angle $\t^\ast$ strongly deviates from the quasi-classical
value $0$ and even comes close to $\pi/2$, the device is then mainly
supplied by the measurement apparatus rather than the heat bath.
In fact, Jacobs argued that, for thermodynamical consistency,
this type of energy input must be considered as work rather than heat
\cite{jaco09}.
It should, however, also be clearly distinguished from the
work output extracted by external control operations.
Further clarification of the role of the measurement process in this
context,using e.g. a scheme proposed in \cite{alla13a}, constitutes an
important and challenging subject for future research.

\section*{References}
\addcontentsline{toc}{section}{References}
\providecommand{\href}[2]{#2}
\begingroup
\raggedright
\endgroup

\end{document}